\documentclass[conference]{IEEEtran}
\IEEEoverridecommandlockouts
\usepackage{amsmath,amssymb,amsfonts}
\usepackage{algorithmic}
\usepackage{graphicx}
\usepackage{bbm}
\usepackage{textcomp}
\usepackage{xcolor}
\def\BibTeX{{\rm B\kern-.05em{\sc i\kern-.025em b}\kern-.08em
    T\kern-.1667em\lower.7ex\hbox{E}\kern-.125emX}}

\newcommand\Tstrut{\rule{0pt}{2.6ex}}         

\begin{document}

\title{A Data-driven Nonlinear Recharge Controller for Energy Storage in Frequency Regulation}

\author{\IEEEauthorblockN{Wenting Ma}
\IEEEauthorblockA{{Dept. of Electrical Engineering} \\
{Tsinghua University}\\
Beijing, China \\
mwt17@mails.tsinghua.edu.cn}
\and
\IEEEauthorblockN{Bolun Xu}
\IEEEauthorblockA{{Dept. of Earth and Environmental Engineering} \\
{Columbia University}\\
New York, NY, USA \\
bx2177@columbia.edu}

\thanks{Wenting Ma was a visiting student at Columbia University in Spring 2020 supported by the Columbia College Visiting Student Program and Tsinghua University's International Class of Energy Internet Program.}
}

\maketitle

\begin{abstract}
Battery energy storage boosts up the response speed of power system frequency regulation, but must be recharged carefully to minimize the distortion to the frequency regulation response. This paper proposes a nonlinear feedback controller to optimize the recharge for storage resources in frequency regulation. This controller is designed using a data-driven best-hindsight optimization framework, the resulting nonlinear recharge controller's gain depends on the storage state of charge as well as its power and energy rating. The developed controller is compared with two benchmark automatic generation control designs, one is a proportional-integral-based control from PJM Interconnection, the other one is based on linear-quadratic regulator. Simulation results using real area control error data from PJM Interconnection show the proposed controller achieves smaller deviations in both the area control error and the  storage state of charge compared to the two benchmark controllers under various storage configurations. 
\end{abstract}


\section{Introduction}
Frequency regulation was one of the first grid-scale applications for battery energy storage  (BES) and was considered an ideal BES application due to its high requirement of response speed and low requirement of energy capacity. Power system frequency stability depends on the balance between generations and demands, and is critical to the reliable supply of electric energy. Units participating in frequency regulation by adjusting their active power output according to automatic generation control (AGC) signals instructed by the system operator. With the increasing penetration of intermittent renewable generations, conventional thermal units can hardly follow the disturbance dynamics resulting from the loss of rotational inertia and increasing time-variant generation~\cite{ulbig2014impact}, and BES becomes an ideal choice to boost-up the response speed regulation services and reduce the capacity procurement requirement~\cite{makarov2008assessing}. 

Yet, a new challenge arises in systems where BES participants are taking up a significant market share in frequency regulation services: \emph{to manage the recharge need of these BES while minimizing the distortion to the regulation response}. PJM Interconnection, the largest system operator in North America, was a pioneer in integrating BES and its regulation market share of BES peaked 50\% in 2018~\cite{xu2018optimal}. These BES follow a filtered AGC signal (RegD signal) to make sure the net energy consumption is less than 15 minutes (normalized by the rated power of the BES), so that BES with limited energy capacity can be utilized more effectively than through the traditional regulation signal (RegA signal)~\cite{wang2019battery}. However, PJM has experienced cases that the recharge need for these storage out-weights the AGC, or these BES have no more energy left when the system is still experiencing a contingency. As a result, a higher BES ratio can worsen the regulation quality and weaken the system reliability, and PJM has revised its AGC design and enforced a 40\% cap on storage participation~\cite{pjm_agc}.


The key challenge in designing AGC for storage is to incorporate its charge/discharge efficiency and tight energy limits, which are nonlinear dynamics that cannot be well addressed by common linear system controllers such as proportional-integral (PI) or linear-quadratic regulator (LQR)~\cite{isidori2013nonlinear}.  On the other hand, frequency fluctuations in power systems are non-Gaussian due to human interactions and renewable intermittency~\cite{schafer2018non}, which also degrades the performance of traditional LQR controllers, while emerging researches on data-driven AGC have not considered the recharge need for storage resources~\cite{hidalgo2019frequency}. The novelty of this paper is the design of a controller that learns the optimal control policies from historical regulation signals without assuming any underlying distributions and actively incorporates storage constraints. The policy is implemented as a lookup table and can be seamlessly incorporated into existing PI-based AGC designs for managing storage recharge, while the rest of the AGC, especially the control loop for conventional generators, remains unchanged. The main contributions of this paper can be summarized as follows:
\begin{itemize}
\item We design a nonlinear feedback controller to manage storage state of charge (SoC) in frequency regulation using a data-driven best-hindsight optimization~\cite{hazan2020nonstochastic} approach.
\item We conclude best-hindsight solutions into a feedback gain lookup table and incorporate it with AGC for conventional generators.
\item We tested the performance of the controller against benchmark AGCs under different BES configurations using historical PJM regulation signals.
\end{itemize}

The rest of the paper is organized as follows: Section \ref{PF} presents the problem formulation and introduces benchmark controllers, Section \ref{ECFR} describes the best-hindsight optimization problem and policy selection of proposed BES controller. Section \ref{CS} evaluates the performance of the controller by simulation, and performs a quantitative correlation analysis between initial battery states and controller gains, and finally section \ref{C} concludes with our findings.

\section{Formulation and Benchmarks}\label{PF}
We start by introducing the set-up for the frequency regulation simulation environment and unit response models, then introduce the two benchmark AGC designs. We also introduce an anti-windup design into the AGC to adapt linear controllers for handling tightly constrained BES units.

\subsection{Simulation Framework}
\begin{figure}[htbp]
\centerline{\includegraphics[width = .75\columnwidth]{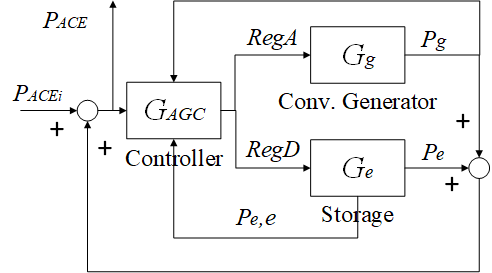}}
\caption{The AGC block diagram.}
\label{flowchart}
\end{figure}

To compare the performance of different controllers, we first set up a frequency regulation simulation environment according to the PJM AGC design manual~\cite{pjm_agc}. Fig.~\ref{flowchart} illustrates the flowchart of this model. Signal input into this simulation is the uncorrected area control error (ACE) signal, which can be reconstructed using the historical PJM regulation signals, unit capacity, and the unit response model (See Section~\ref{sec:model}). The ACE signal is the sum of the corrected ACE signal plus the response from all regulation units, and it is feed into the AGC and generates RegA and RegD signals to conventional generators and BES participants. Responses from all regulation units are added to the ACE. Note that this framework is similar to signal tracking and does not consider the frequency dynamics, because in U.S. frequency regulation belongs to the secondary frequency control tier and its propose is to mitigate ACE and relieve the primary response, instead of stabilizing system dynamics~\cite{xu2016comparison}. 

\subsection{Modeling of the Conventional Generators and BES}~\label{sec:model}
\begin{figure}[htbp]
\centerline{\includegraphics[scale = 0.45]{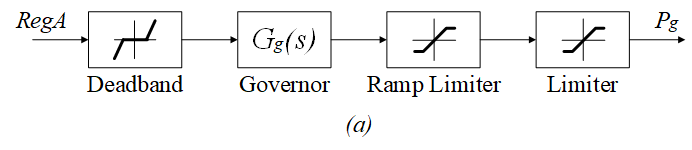}}
\centerline{\includegraphics[scale = 0.45]{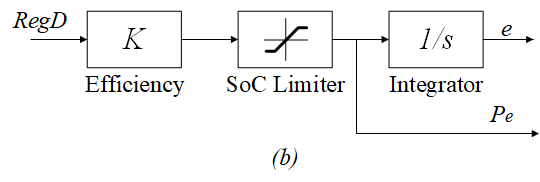}}
\caption{(a) Conventional generator model. (b) BES model}
\label{generator}
\end{figure}

Conventional generator response model~\cite{doenges2019improving} is described in Fig.~\ref{generator}~(a). After the controller sends RegA signal to generators, a deadband is applied to prevent excessively frequent response. $G_g(s)=1/(T_gs+1)$ is the governor dynamics. Power output of generators $P_g$ is constrained by ramp rates and maximum generation capacity.

The response model of BES is described in Fig.~\ref{generator}~(b), which considers charging and discharging efficiency as well as SoC constraints. $e$ is the SoC of BES, and is fed back to the controller.

\subsection{PJM Conditional Neutrality Controller}

Beginning on Jan.9, 2017, the PJM's new conditional neutrality AGC~\cite{pjm_agc} was put into production. It is a hybrid proportional-integral (PI) controller including a RegD integral feedback loop  so that RegD is energy neutral. This AGC sends slow RegA signal to conventional generators with limited ramping capability and unlimited energy, and send the remaining RegD signal to faster resources with limited energy\cite{wang2019battery}.

\begin{figure}[htbp]
\centerline{\includegraphics[scale=0.38]{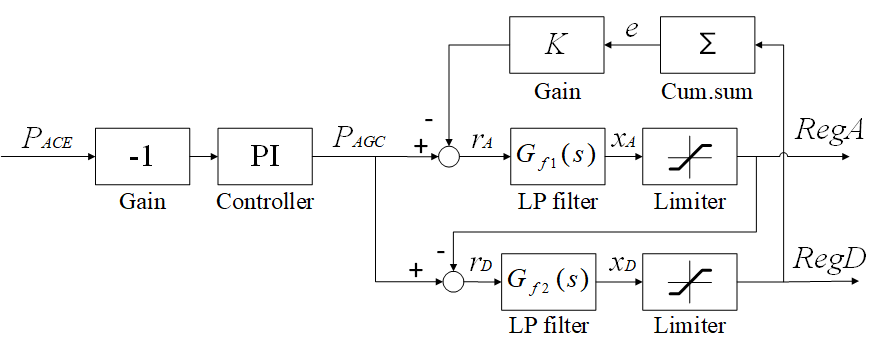}}
\caption{Structure of PJM Conditional Neutrality Controller.}
\label{PJMcontroller}
\end{figure}

The structure of this controller is shown in Fig.~\ref{PJMcontroller}, where $P_{ACE}$ denotes filtered ACE, $G_{f1}(s)=1/(T_as+1)$ is the low pass filter for RegA signal, and $G_{f2}(s)=1/(T_ds+1)$ is the low pass filter for RegD signal. The PI control implemented in this controller is given as the following:
\begin{subequations}
\begin{align}
P_{AGC} &= -K_P P_{ACE}-K_I I_{ACE} \label{eq1} \\
I_{ACE} &=\int P_{ACE} dt\label{eq5} 
\end{align}
\end{subequations}
$P_{AGC}$ is the signal output of PI controller, $K_P$ and $K_I$ denotes controller gains of the  proportional term and integral term. Parameters of limiters are set according to output power constraints on  RegA and RegD resources within the area.

\subsection{LQR-based AGC}
The PJM controller determines the RegD signal as the residual between the actual AGC and the low-passed RegA signal. This prohibits adjustments of the control policy to further minimize corrected ACE ($P_{ACE}$ in Fig.~\ref{flowchart}) for a linear time-invariant system. Besides, a regulating action performed by BES suffers efficiency losses so that storage will slowly become empty. We introduce another benchmark AGC based on LQR to determine the trade-off between AGC performance and regulation costs. Although the use of LQR in designing AGC is questionable because the regulation system is nonlinear and ACE disturbances are not Gaussian, our goal is to provide another benchmark in the comparison, and this LQR-based AGC does show competitive performance compared to the PJM AGC in simulation. 

The continues time-domain LQR AGC formulation is 
\begin{subequations}
\begin{align}
\min_{x,u} \; &J := \int_{t_0}^{\infty} x^TQx+u^TRu\label{eq2a}\\
&\dot{x}=Ax+Bu+\omega \label{eq2c}
\end{align}\label{eq2}
The LQR AGC uses a pre-set PI control for RegA and optimizes the RegD control, the states include ACE, RegA, integral of ACE, and BES SoC. The control is the RegD signal.
\begin{align}
x=&[P_{ACE},RegA,I_{ACE},e]^T\label{eq7}\\
u_t=&RegD
\end{align}
Based on the modeling approach of system dynamics proposed in \cite{liu2012enhanced}, the state and input matrices $A$ and $B$ can be given as 
\begin{align}
A=&\begin{bmatrix}-\frac{1}{M}&\frac{1}{M}&0&0\\
			-\frac{K_P}{T_a}&-\frac{1}{T_a}&-\frac{1}{T_a}&0\\
			K_I&0&0&0\\
			0&0&0&0\end{bmatrix}\label{eq8}\\
B=&\begin{bmatrix}\frac{1}{M}&0&0&1\end{bmatrix}\label{eq9}
\end{align}
where $M$ is the normalized inertia constant. The LQR generates a state-feedback control for RegD as
\begin{equation}
u=-Kx\label{eq4}
\end{equation}
where $K$ is the regulation control gain of system states.


\end{subequations}

\subsection{Anti-windup Control}
We add anti-windup control~\cite{zheng1994anti} to all AGC designs to avoid integral wind-up in the integrals caused by control saturation. Nonentities in regulation units, including governor dead-bands, ramp rates, and storage energy constraints, introduces nonlinear dynamics and sometimes even cause instability~\cite{tan2017load}. When the actuator saturates as reached a constraint, the integral term in \eqref{eq1} tends to accumulate the control error and dominate the controller, so called \emph{wind-up}, and a return to normal control operations becomes difficult due to the large integral value. Wind-up effects are more likely to occur in systems with high BES share as storage may frequently reach upper or lower energy limits~\cite{xu2018optimal}, though they are less noticeable in traditional AGC because conventional generators are less likely to be constrained,  

We modify the integrator in \eqref{eq5} by adding a comparing term of the actual input and commanded input. As shown in  Fig.~\ref{flowchart}, we suppose certain states of actuators can be measured and fed back to the controller. This includes the actual generators output $P_g$, BES power output $P_e$, and SoC of BES $e$. The integrator is now written as:
\begin{equation}
I_{ACE}=\int[P_{ACE}+(RegA-P_g)+(RegD-P_e)]dt\label{eq10}
\end{equation}
The scheme is not in effect when system's actual output is synchronized with the control signal, but feeds back the mismatch when non-linearities are encountered.

\section{Optimization-aided Controller Design}\label{ECFR}
We introduce a data-driven best-hindsight optimization problem based on the existing PJM AGC to derive the optimal recharge policy for BES given their nonlinear dynamics, and implement this policy as a look-up table. Our storage control (RegD) is defined as a PI control combined with a nonlinear SoC feedback control as
\begin{align}
    RegD_t = -K^D_p P_{ACE,t} - K^D_I I_{ACE,t} - f_{SoC}(e_t)e_t\label{eq11}
\end{align}
where $K^D_p$ and $K^D_I$ are proportional and integral control gains for the RegD signal in $MW$, while $f_{SoC}(e_t)$ is the SoC feedback control gain and it is a function dependent on the storage state-of-charge value $e_t$ in $MW\cdot h$. 

We choose the SoC feedback gain as the only nonlinear control component based on the observation that during the design process the PI gain does not show meaningful dependency on the storage SoC. Also, system operators usually have pretty good knowledge about the choice of PI gains for their AGC but are less familiar with how to manage storage SoC, therefore this study focused on the SoC feedback gain and assume it to be a scalar function to the storage SoC.


\subsection{Best-hindsight Optimization Formulation}
We use a best-hindsight  framework to design the feedback gain for state-of-charge $f_{soc}(e_t)$ in order to better incorporate the BES non-linearity due to efficiency and energy limits subjects to the non-Gaussian ACE signal with unknown distribution ($p<0.001$ in both Jarque-Bera test and Lilliefors test). In the best-hindsight framework, we solve for the optimal SoC feedback gain $K_e$ over a relatively short period $T$ with a starting time index $t_0$ using historical ACE data as a deterministic optimization problem. We repeatedly solve this optimization for various $t_0$ and initial SoC $e_0$, and record all optimized SoC feedback gains ($K_e$), which we eventually concluded it into a function with respect to the storage state-of-charge. Such that the control policy is data-driven as it depends on the historical realizations of the ACE. The formulation for this deterministic optimization problem is

\begin{subequations}
\begin{align}
\min_{K_{e}} J&=\Sigma_{t=t_0}^{t_0+T}[(P_{ACE,t})^2+w_e\cdot e_t^2]\label{eq12a}\\
\text{s.t. }& \nonumber\\
RegA_t &= -K_p P_{ACE,t} - K_I I_{ACE,t}\label{eq12e} \\
RegD_t &= -K^D_p P_{ACE,t} - K^D_I I_{ACE,t} - K_e e_t \label{eq12f} \\
& \text{RegD control as in \eqref{eq11}} \nonumber\\
&\text{system and unit response models as in Fig.~\ref{flowchart} \& \ref{generator}}\nonumber
\end{align}\label{eq12}
\end{subequations}

This  program optimizes a SoC feedback gain $K_e$ that minimizes the objective using given PI control settings. \eqref{eq12a} is the cost function, where $w_e$ is the weight coefficient on its SoC deviation term. By modifying the value of $w_e$, the relative importance between reducing output ACE and keeping SoC close to its desired state is adjusted. \eqref{eq12e} calculates the RegA signal using a pre-set PI control setting, and \eqref{eq12f} calculates the RegD signal using a different PI control setting and a SoC feedback gain $K_e$, which is the decision variable in this problem. The two regulation signals are then sent into unit response models as shown in Fig.~\ref{generator} according to the simulation framework in Fig.~\ref{flowchart}.


\subsection{Solution Method}
We obtain the feedback gain function $f_{SoC}$ by fitting results of $K_e$ in \eqref{eq12} to a uni-variate function depending on storage SoC $e$. \eqref{eq12} is a nonlinear optimization problem because of the quadratic objective and the use of saturation functions, and it is solved using the {\fontfamily{qcr}\selectfont fmincon} function from MATLAB.

We train $f_{SoC}$ using historical ACE signal $w_t$ over a period $t\in\{1,2,\dotsc, H\}$ with following steps. 

\begin{enumerate}
\item Discretize value range of SoC $e\in[0,E]$ into $N=500$ bins: $\{e_n|n\in\{1,2,\dotsc, N\}\}$.
\item \textbf{For} $t_0 = 1:(H-T)$ 

Solve \eqref{eq12}, record the solution $K_e$ and its corresponding initial SoC level $e$: $K_e(t_0)$, $e(t_0)$.
\item $f_{SoC}(e)$ is obtained by computing the average value of recorded $K_e$s:
\begin{equation}
f_{SoC}(e)=\frac{\sum_{t_0=1}^{H-T}K_e(t_0)\mathbbm{1}\{e_{n-1}\leq e(t_0)<e_n\}}{\sum_{t_0=1}^{H-T}\mathbbm{1}\{e_{n-1}\leq e(t_0)<e_n\}}\label{eq13}
\end{equation}

\end{enumerate}

\subsection{Training Results and Insights}
\begin{figure}[t]
\centering 
\includegraphics[width = .85\columnwidth, trim = 0mm 0mm 0mm 0mm, clip]{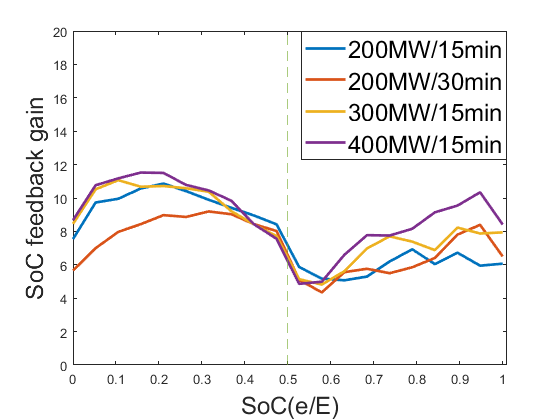}
\caption{Averaged recharge control policy function $f_{SoC}(e)$ under various BES configurations. The feedback gain is applied to the SoC deviation to the reference $50\% $ target in MWh units.}
\label{Kvalues}
\vspace{-.5cm}
\end{figure}

We fix RegA capacity to $C_a=400MW$ according to PJM regulation requirements~\cite{pjm_agc}. We train the lookup table $f_{SoC}(e)$ under various BES configurations. The obtained feedback gain function $f_{SoC}(e)$ is nonlinear, as visualized Fig.~\ref{Kvalues}. Here, a larger BES energy capacity is represented by a longer discharge time. Feedback gains from different BES settings share some common features, such as the gain is at minimal when the SoC is around the reference 50\% level, and increases as the SoC approaches storage limits. An interesting result is that the feedback gain is reduced when BES is nearly empty (SoC less than 10\%), which is likely due to that the system is experiencing contingencies and has a persistent demand for regulation generation that depletes the storage. A large recharge power in this scenario may add too much stress to the AGC and worsen the overall regulation quality, which is a similar concept to PJM's current practice of manually disabling the energy neutrality control during contingencies. 

\section{Simulation and Results}\label{CS}
We present simulation results of the proposed controller and benchmark controllers using real uncorrected ACE data from PJM, i.e., the ACE of the system assuming there is no regulation response. The uncorrected ACE data is reconstructed using historical regulation signals and unit response models according to the PJM AGC manual~\cite{pjm_agc}. 30 days of 2-second-resolution data are used to construct the proposed recharge policy, and the controller performance is evaluated using another 15 days of data. All AGC have the same setting of PI control gains for conventional regulation units (RegA) with $K_P=0$, $K_I=0.4$. The rest of the benchmark PJM controller parameters are based on the PJM manual~\cite{pjm_agc}. The benchmark LQR controller is tuned to achieve the best control performance using cost weights $Q(1,1) = 1$, $Q(2,2) = (C_d/C_a)^2$, $Q(3,3) = 0$, $Q(4,4) = 1/E^2$, and $R = 1$. The resulting state-feedback control gain $K$ is $K=[0.4309,	-0.0339,0.1210,8]$. {In our simulations, the LQR control performance is not sensitive to different storage settings so the same $K$ is used for all storage cases.} For the proposed controller, we set the RegD PI control setting to $K^D_P=1$ and $K^D_I=0.8$, the window length $T$ for best-hindsight optimization in \eqref{eq12a} to 15 minutes. {The SoC deviation weight $w_e$ is tuned using an exhaustive search in each storage setting to achieve the best overall control performance}. 

\begin{figure}[t]
\centerline{\includegraphics[width = .95\columnwidth, trim = 5mm 5mm 10mm 0mm, clip]{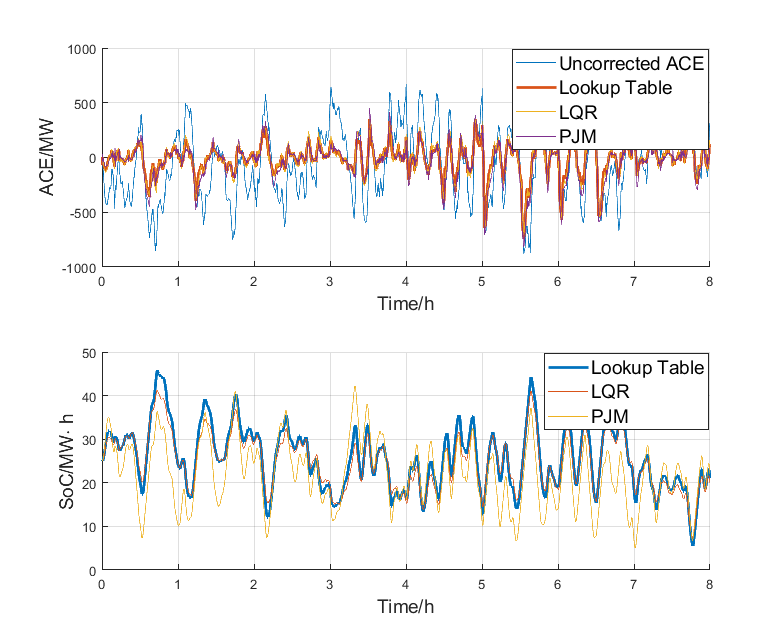}}
\caption{Comparison of controller performance. Overall, the three controllers provide similar shapes of ACE control results and SoC profiles, while the proposed control achieves the smallest ACE deviations and SoC deviations. }
\label{performance}
\end{figure}

We simulate 400MW RegA capacities (i.e. conventional generators) with varying RegD capacities as well as the BES energy capacity. {We assume all RegD is provided only by BES, and assume all BES units have identical physical parameters, initial SoC, and response characteristic, so all BES units can be aggregated as a single BES in the design and simulations.} The BES round-trip efficiency is 85\%. Fig.~\ref{performance} demonstrates comparisons of the uncorrected ACE, responses from different AGC designs, and the resulting total SoC from all storage.  

\begin{table}[htbp]
\caption{Controller Performance Under Different BES Configurations}
\begin{center}
\begin{tabular}{|c|c c c|c c c|}
\hline
\Tstrut
\textbf{BES}&\multicolumn{3}{|c|}{$\overline{(P_{ACE})^2}/\times 10^3$}&\multicolumn{3}{|c|}{$\overline{e^2}$} \\
\cline{2-7} \Tstrut
\textbf{Config.} & Prop. & LQR &  PJM & Prop. & LQR &  PJM\\
\hline\Tstrut
200MW/15min&32.6&34.3&36.5&110.7&101.9&116.1\\
\hline\Tstrut
200MW/20min&31.0&32.4&33.9&152.3&160.1&174.7\\
\hline\Tstrut
200MW/30min&29.5&29.8&30.6&216.4&310.9&323.4\\
\hline\Tstrut
200MW/60min&24.5&26.4&26.3&730.3&1015.7&1008.6\\
\hline\Tstrut
300MW/15min&28.6&29.8&31.3&158.7&206.5&176.7\\
\hline\Tstrut
400MW/15min&26.6&29.8&28.8&189.2&370.4&218.9\\
\hline
\end{tabular}
\label{tab1}
\end{center}
\vspace{-.3cm}
\end{table}

We use two indexes to compare the control performance of different AGC designs: the average values of output ACE square $\overline{(P_{ACE})^2}$ and the SoC deviation square $\overline{e^2}$, where  we set $e_0=50\%$ as our desired state of BES. Table~\ref{tab1} shows the results over the 15 day test period. Note that under the BES configuration of 200MW/60min, we adjust the parameter settings for the proposed controller to $T$=30min and $w_e$ = 10 in order to obtain best performance. The proposed controller is observed to reduce output ACE error by up to 10.74\% compared to LQR controller, and 10.68\% compared to PJM controller. Moreover, it reduces the SoC deviation by up to 48.92\% compared to LQR controller, and 33.09\% compared to PJM controller. This improvement in reducing extreme charge and discharge is especially significant for BES with greater energy capacity such as 200MW/30min and 200WM/60min.

It is worth noting that the performance of the proposed controller depends heavily upon the tuning of SoC deviation weight setting $w_e$. The current setting of $w_e=60$ achieves the overall best performance under a BES configuration of 300MW/15min, where output ACE error is reduced by 4.16\% compared to LQR controller, and SoC deviation is reduced by 10.20\%. Meanwhile, an improvement is expected by tuning $w_e$ under a rather under-performing configuration of BES.

\section{Conclusion}\label{C}
We developed a data-driven nonlinear feedback control to optimize the recharge need for battery energy storage in frequency regulation. This feedback controller uses a best-hindsight framework that models non-linear system dynamics and non-stochastic disturbances, and can incorporate with any existing AGC design. Simulation results using real ACE data shows the proposed controller achieved significant improvements compared to two benchmark AGC in both controlling ACE and preventing excessive use of ESS that might exacerbate degrading. Moreover, it is capable of adapting to different BES configuration by adjusting relative weights accordingly. The correlation between control gains and initial SoC is analyzed to help understand the effect of BES capacity constraints on the controller's choice of control policy.

The key insight from our study is that energy storage recharge can be better managed with nonlinear controllers, which were not considered by any existing AGC design. We hope results from this paper could incentivize new ideas on how to better incorporate storage resources in ancillary services to achieve better grid reliability and sustainability. Future directions includes generalize this control method over diverse storage resources and to large-scale interconnected systems composed of multiple balancing areas. Our results show that new monitoring schemes and market designs are required to improve the utilization for storage resources.

\bibliographystyle{IEEEtran}	
\bibliography{IEEEabrv,literature}		
\end{document}